\documentclass[12pt,draftcls,onecolumn]{IEEEtran}

\usepackage{stmaryrd}
\usepackage{amsfonts}
\usepackage{graphics} % for pdf, bitmapped graphics files
\usepackage{subfigure}
\usepackage{epsfig} % for postscript graphics files
\usepackage{mathptmx} % assumes new font selection scheme installed
\usepackage{times} % assumes new font selection scheme installed
\usepackage{amssymb}  % assumes amsmath package installed
\usepackage{floatrow}
\usepackage{multirow}
\usepackage{lettrine}
\usepackage{algorithm}
\usepackage{algpseudocode}
\usepackage{color}
\usepackage{booktabs}
\usepackage{graphicx}
\usepackage{makecell}
\usepackage{supertabular}
\usepackage{amsmath}
\usepackage{graphicx}    % standard LaTeX graphics tool
\title{\LARGE \bf
Learning Control of Quantum Systems}

\author{Daoyi~Dong% <-this % stops a space
\thanks{D. Dong is with the School of
Engineering and Information Technology, University of New South
Wales, Canberra, ACT 2600,
Australia
        {\tt\small daoyidong@gmail.com}}}%
\begin{document}

%\title{Learning Control of Quantum Systems}

%\author{Daoyi Dong  % more than one author? Use "\and" in between
% \and
% Name of Second Author etc.
%}

% Use the package "url.sty" in the preamble
% to avoid problems with special characters
% used in your e-mail or web address

\maketitle

\begin{abstract}
This paper provides a brief introduction to learning control of quantum systems. In particular, the following aspects are outlined, including gradient-based learning for optimal control of quantum systems, evolutionary computation for learning control of quantum systems, learning-based quantum robust control, and reinforcement learning for quantum control.
\end{abstract}

\section{Introduction}
Controlling quantum systems has become a central task in the development of quantum technologies, and quantum control has witnessed rapid progress in the last two decades; for an overview, see, e.g., the survey papers \cite{Dong-and-Petersen-2010,Rabitz-et-al-2000,Glaser2015,Brif-et-al-2010,Altafini-and-Ticozzi-2012} or the monographs \cite{Wiseman-and-Milburn-2010,D'Alessandro-2007}. The general goal of quantum control is to actively manipulate and control the dynamics of quantum systems for achieving given objectives \cite{NJP-roadmap,Shu-PRL} (e.g., rapid state transfer, high-fidelity gate operation). Two of fundamental issues in quantum control include investigating controllability of quantum systems and designing control laws to achieve expected control systems performance. Controllability is concerned with what control targets can be achieved and the controllability of finite-dimensional closed systems has been well addressed \cite{D'Alessandro-2007}. A few results on the controllability of open quantum systems have also been presented. For control law design, optimal control theory \cite{Dong-and-Petersen-2010}, Lyapunov control approaches \cite{Kuang-et-al-2017}, learning control algorithms \cite{Rabitz-et-al-2000} and robust control methods \cite{Dong2019} have been developed in manipulating quantum systems for achieving various control objectives.

Among various control design approaches, learning control is recognized as a powerful method for many complex quantum control tasks and has achieved great success in laser control of molecules and other applications since the approach was presented in the seminal paper \cite{Judson-and-Rabitz-1992}. Many quantum control tasks may be formulated as an optimization problem and a learning algorithm can be employed to search for an optimal control field satisfying a desired performance condition. Gradient algorithms have been demonstrated to be an excellent candidate for numerically finding an optimal field and have achieved successful applications in nuclear magnetic resonance (NMR) systems due to their high efficiency \cite{Khaneja-et-al-2005}. In many other optimal control problems, the gradient information may not be easy to obtain and some complex problems may have local optima. For these situations, stochastic search algorithms usually have improved performance to find a good control field. The genetic algorithm (GA) and differential evolution (DE) \cite{Dong2019} have been widely used in the area of quantum control of molecular systems and achieved great success \cite{Rabitz-et-al-2000}. Another task in quantum control is to achieve robustness performance in quantum systems. Gradient-based learning algorithms and stochastic search algorithms may be smartly modified to search for robust control fields. Also, some other machine learning algorithms such as reinforcement learning have found successful applications in various tasks (e.g., quantum error correction \cite{DRL-PRX}).

\section{Gradient-based learning for optimal control of quantum systems}
Consider a finite-dimensional quantum control system where its state $|\psi(t)\rangle$ (using the Dirac notation) is described by the following Schr\"{o}dinger equation (setting $\hbar=1$):
\begin{equation} \label{systemmodel}
  \frac{d}{dt}|{\psi}(t)\rangle=-i[H_{0}+\sum_{m=1}^{M}u_{m}(t)H_{m}]|\psi(t)\rangle,\quad t\in [0, T],
\end{equation}
where $H_{0}$ is the free Hamiltonian of the system and
$H_{c}(t)=\sum_{m=1}^{M}u_{m}(t)H_{m}$ is the control Hamiltonian at
time $t$ that represents the interaction of the system with the
external fields $u_{m}(t)$. The $H_{m}$
are Hermitian operators through which the controls couple to the
system. The objective of quantum optimal control is to find control fields $u_{m}(t)$ for maximizing a performance functional $\Phi$. $\Phi$ may be a given functional of the state $|\psi\rangle$ and control defined according to practical requirement. For example, the fidelity $\Phi=|\langle\psi(T)|\psi_f\rangle|^{2}$ between the final state $|\psi(T)\rangle$ and target state $|\psi_f\rangle$ or the expectation $\Phi=|\langle\psi(T)|\hat{O}|\psi(T)\rangle|^{2}$ of an operator $\hat{O}$ may be defined as a performance index for state transfer task. In order to maximize the performance $\Phi$, we may employ the GRAPE (gradient ascent pulse engineering) algorithm \cite{Khaneja-et-al-2005} or the Krotov method \cite{Krotov-method2014} to search for the control field. For simplicity, we may discretize time $T$ in $N$ equal steps and during each step let the control fields $u_{m}$ be constant. The basic idea in the GRAPE algorithm is that the control fields are iteratively updated following the gradient direction of $\frac{\delta \Phi}{\delta u_{m}(k)}$ with a learning rate $\eta$, i.e.,
\begin{equation}
u_{m}(k+1)=u_{m}(k)+\eta \frac{\delta \Phi}{\delta u_{m}(k)}.
\end{equation}

The gradient-based learning method can also be extended to the optimal control problem of unitary transformations (e.g., quantum gates) and open quantum systems. For example, for a unitary transformation $U$, its evolution is described by the following equation
\begin{equation}
  \dot{U}(t)=-i[H_{0}+\sum_{m=1}^{M}u_{m}(t)H_{m}]U(t), \ \ \ \ U(0)=I.
\end{equation}
Now the objective
is to design the controls $u_{m}(t)$ to steer the unitary $U(t)$ from $U(0)=I$ to a desired target
$U_{F}$ with high fidelity. We may define the performance function as
$\Phi=|{\langle U_F|e^{i\varphi}U(T)\rangle}|^2$ for an arbitrary phase factor $\varphi$. Then we can calculate the gradient
$\delta\Phi/\delta u_m(k)$, and the optimal control field can be searched for by following the gradient \cite{Dong-2016SR}.
When we consider the optimal control problem of an open quantum system, its state should be represented by a density matrix $\rho$ and its dynamics should be described by a master equation. The dynamics of a Markovian open quantum system can be described using the following master equation in the Lindblad form as \cite{Wiseman-and-Milburn-2010}
\begin{equation}\label{eq:markovian equations}
         \dot{\rho}(t)=-i[H_{0}+\sum_{m=1}^{M}u_{m}(t)H_{m},\rho(t)]+\sum_{k}\gamma_k\mathcal{D}[L_k]\rho(t),
\end{equation}
where the non-negative coefficients $\gamma_k$ specify the relevant relaxation rates, $L_k$ are appropriate Lindblad operators and $\mathcal{D}[L_k]\rho=(L_k \rho L_k^{\dagger}-\frac{1}{2} L_k^{\dagger} L_k \rho-\frac{1}{2}\rho L_k^{\dagger} L_k ).$
The open GRAPE algorithm has also been developed to calculate the gradient based on the master equation (see \cite{OpenGrape4-Glaser-2011}).

Using the basic idea of gradient-based learning control, some variants have been developed for various requirements in quantum optimal control. For example, a data-driven gradient optimization algorithm (d-GRAPE) has been proposed to correct deterministic gate errors in high-precision quantum control by jointly learning from a design model and the experimental data from quantum tomography \cite{Wu-dGRAPE}. A gradient-based frequency-domain optimization algorithm has been developed to solve the optimal control problem with constraints in the frequency domain \cite{Shu-PRA2016}. Existing results show that gradient-based learning methods can usually achieve excellent performance for solving optimal control problems when the system model is known and the dynamics can be equivalently (or approximately) described using a closed quantum system. This is also analyzed using quantum control landscape theory \cite{Chakrabarti-and-Rabitz-2007}.

\section{Evolutionary computation for learning control of quantum systems}
Gradient algorithms have shown powerful capability for numerically finding optimal controls due to their excellent performance \cite{Khaneja-et-al-2005}. In many practical applications, it may be difficult to obtain the gradient information or there exist local optima in complex quantum control problems. For these situations, a natural idea is to employ stochastic search algorithms to seek good controls. Evolutionary computation including GA and DE has been widely used in the area of quantum control. In these evolutionary computation methods, crossover, mutation and selection operations are iteratively implemented to search for good solutions (optimal controls) in a parameter space. For example, a subspace-selective self-adaptive differential evolution (SUSSADE) algorithm has been proposed to achieve a high-fidelity single-shot Toffoli gate and single-shot three-qubit gates \cite{Zahedinejad-et-al-2015}, \cite{Zahedinejad-et-al-2016}. Existing results showed that DE with equally-mixed strategies can achieve improved performance for quantum control problems \cite{Ma2017CTT}. Several promising evolution algorithms have been investigated comparatively in \cite{Zahedinejad-et-al-2014} and it was found that DE usually outperformed GA and particle swarm optimization for hard quantum control problems.

The above introduction of gradient-based learning and evolutionary computation mainly involves open-loop control strategies. Evolutionary computation has demonstrated extremely powerful capability when it is integrated into closed-loop control design. Closed-loop learning control, where each cycle of the closed-loop is executed on a new sample, has achieved great successes in the
laser control of laboratory chemical reactions \cite{Rabitz-et-al-2000,Judson-and-Rabitz-1992}. A closed-loop learning control procedure generally involves three components \cite{Dong-and-Petersen-2010,Rabitz-et-al-2000}: (i) a trial laser control input, (ii) the laboratory generation of the control that is applied to the sample and subsequently observed for its impact, and (iii) a learning algorithm to suggest the form of the next control input by considering the prior experiments. The initial trial control input may be a random input field or a well-designed laser pulse. A feature of a good closed-loop learning control design is its insensitivity to the initial trials. A key task is to develop a good learning algorithm for ensuring that the learning process converges to achieve a predetermined objective. GA, DE and several rapid convergence algorithms have been developed for this task \cite{Brif-et-al-2010,Judson-and-Rabitz-1992}. The optimal control problem is usually formulated as solving an optimization problem by maximizing a functional which is related to some variables such as the control inputs, quantum states and control time but may have no analytical form. In the learning process, the optimization problem is solved iteratively. First, a trial input is applied to a sample to be controlled and the result is observed. Second, a learning algorithm suggests a better control input based on the prior experiments. Third, the ``better" control input is applied to a new sample. This process continues until the control objective is achieved or the maximum permitted iteration number is reached. It is often feasible to produce many identical-state samples for laboratory chemical molecules. If the control objective is well selected, there is a capability to apply specified control inputs to the samples, and the learning algorithm is sufficiently smart for searching for good control inputs, this process will converge and an optimal control pulse can be
found \cite{Rabitz-et-al-2000}.

\section{Learning-based quantum robust control}
The robust control of quantum systems has been recognized as a key task in developing practical quantum technology since the existence of noise and uncertainties is unavoidable. Learning control is an effective candidate for achieving robust performance in some quantum control problems \cite{Dong2019}. We first consider the control problem of inhomogeneous quantum ensembles. An inhomogeneous quantum ensemble consists of many individual quantum systems (e.g., atoms, molecules or spin systems) and the parameters describing the system dynamics of these individual systems could have variations \cite{Li-and-Khaneja-2006,Chen-et-al-2014}. An example is that a spin ensemble in NMR may encounter large dispersion in the strength of the applied radio frequency field and there also exist variations in the natural frequencies of these spins \cite{Li-and-Khaneja-2006}. Inhomogeneous quantum ensembles have wide applications in many fields ranging from quantum memory to magnetic-resonance imaging. Hence, it is highly desirable to design control laws for an inhomogeneous ensemble to employ the same control inputs to steer individual systems with different dynamics from a given initial state to a target state.

A sampling-based learning control (SLC) method has been developed to achieve high fidelity control of inhomogeneous quantum ensembles \cite{Chen-et-al-2014}.
Consider an inhomogeneous ensemble in which the Hamiltonian of each individual system has the following form
\begin{equation}\label{inhomogeneousHamiltonian}
H_{\omega, \theta}(t)=\omega H_{0}+\sum_{m=1}^{M}\theta u_{m}(t)H_{m}.
\end{equation}
We assume that the parameters $\omega\in [1-\Omega, 1+\Omega]$ and $\theta \in [1-\Theta, 1+\Theta]$, and the constants $\Omega \in [0,1]$ and $\Theta \in [0,1]$ represent the
bounds of the parameter dispersion. The objective
is to design the controls $\{u_{m}(t)\}$ to simultaneously stabilize the
individual systems (with different $\omega$ and $\theta$) of the quantum
ensemble from an initial state $|\psi_{0}\rangle$ to the same target state
$|\psi_{f}\rangle$ with high fidelity.
This task can be achieved using the SLC method including two steps
of ``training" and ``testing and evaluation" \cite{Chen-et-al-2014}. In the training step, we select $N$ samples from the quantum
ensemble regarding the distribution (e.g., uniform
distribution) of the inhomogeneity parameters and then construct a
generalized system as follows
\begin{equation}\label{generalized-system}
\frac{d}{dt}\left(%
\begin{array}{c}
  |{\psi}_{\omega_1,\theta_1}(t)\rangle \\
  |{\psi}_{\omega_2,\theta_2}(t)\rangle \\
  \vdots \\
  |{\psi}_{\omega_N,\theta_N}(t)\rangle \\
\end{array}%
\right)
=-i\left(%
\begin{array}{c}
  H_{\omega_1,\theta_1}(t)|\psi_{\omega_1,\theta_1}(t)\rangle \\
  H_{\omega_2,\theta_2}(t)|\psi_{\omega_2,\theta_2}(t)\rangle \\
  \vdots \\
  H_{\omega_N,\theta_N}(t)|\psi_{\omega_N,\theta_N}(t)\rangle \\
\end{array}%
\right)
\end{equation}
where $H_{\omega_n,\theta_n}=\omega_{n} H_{0}+\sum_{m} \theta_{n} u_{m}(t)H_{m}$
with $n=1,2,\dots,N$. The cost function for the generalized system
is defined by
\begin{equation}\label{eq:cost}
\Phi_N(u):=\frac{1}{N}\sum_{n=1}^N \Phi_{\omega_n,\theta_n}(u).
\end{equation}
The task of the training step is to find a control
strategy $u^*$ to maximize the cost functional $\Phi_N(u)$. A gradient-based learning algorithm (s-GRAPE) can be developed to complete this task.
In the process of testing and evaluation, a number of
sampling individual systems are randomly selected to evaluate the control
performance. Results show that the SLC method is potential for control design of various inhomogeneous quantum ensembles (including inhomogeneous open quantum ensembles).

Besides inhomogeneous quantum ensembles, the SLC method is useful for robust control of single quantum systems with various uncertainties. For example, Eq. (\ref{inhomogeneousHamiltonian}) can also correspond to the Hamiltonian of a quantum system with inaccurate model parameter $\omega$ and uncertain multiplicative noise $\theta$. In order to achieve robust control for such a quantum system, we may employ the SLC method to search for robust control pulses \cite{Dong-et-al-2015,Dong-2015SR,Wu-et-al-2017}. The performance of SLC approach can be further improved by exploring the richness and diversity of samples. Inspired by deep learning, a batch-based gradient algorithm (b-GRAPE) has been presented for efficiently seeking robust quantum controls, and numerical results showed that b-GRAPE can achieve improved performance over the SLC method for remarkably enhancing the control robustness while maintaining high fidelity \cite{WuR-PRA-2019}. In other applications where we need to enhance the robustness in closed-loop learning control, we may either use the Hessian matrix information \cite{Xing-NJP} or integrate the idea of SLC into the learning algorithm in searching for robust control fields. For example, an improved DE algorithm (called as \emph{msMS}\_DE) has been proposed to search for robust femtosecond laser pulses to control fragmentation of the molecule $\text{CH}_2\text{BrI}$ \cite{Dong2019}. In \emph{msMS}\_DE, multiple samples are used for fitness evaluation and a mixed strategy is employed for the mutation operation.

\section{Reinforcement learning for quantum control}
Reinforcement learning (RL) \cite{Sutton-and-Barto-1998} is another important machine learning approach and it addresses the
problem of how an active agent can learn to approximate an optimal
strategy while interacting with its environment. It is a model-free feedback-based approach and works well even when the system model is unknown or with uncertainties. RL has been used for learning control of quantum systems. For example, a fidelity-based probabilistic Q-learning approach has been presented to naturally solve the balance problem between exploration and exploitation and was applied to learning control of quantum systems \cite{Chen-et-al-2012TNNLS}. The authors in \cite{RL-PRX} showed that the performance of RL is comparable to optimal control approaches in the task of finding a short and high-fidelity protocol, controlling from an initial to a given target state in nonintegrable many-body quantum systems of interacting qubits. RL can also help identify variational protocols with nearly optimal fidelity even in the glassy
phase. In \cite{Liu-npj}, deep reinforcement learning is employed to simultaneously optimize the speed and fidelity of quantum computation against both leakage and stochastic control errors. A universal quantum control framework was presented to improve the control robustness by adding control noise into training environments for RL agents trained with trusted-region-policy-optimization. By utilizing two-stage learning with teacher and student networks
and a reward quantifying the capability to recover the quantum information stored in a quantum system, the authors in \cite{DRL-PRX} showed how a network-based ``agent" in RL can discover good quantum-error-correction strategies to protect qubits against noise.

\section{Conclusions}
Machine learning has shown powerful capability in discovering high quality controls to achieve optimal control and enhance robust performance for quantum systems. On one hand, it is necessary to further develop or improve existing machine learning algorithms to effectively solve complex quantum control problems emerged from new quantum technologies. On the other hand, various cutting-edge machine learning techniques should be able to find new application opportunities in the area of quantum control.

%
% Non-BibTeX users please use

%%%%%%%%%%%%%%%%%%%%%%%%%%%%%%%%%%%%%%%%%%%%%%%%%%%%%%%%%%%%%%%%%%%%%%

\end{document}